\documentclass[showpacs,prd]{revtex4}
\usepackage{amsmath}
\usepackage{graphicx}
\usepackage{amssymb}
\usepackage{mathpazo}
\usepackage{float}
\usepackage{dcolumn}

\newcommand{\ignore}[1]{}

\newcommand{\be}{\begin{equation}}
\newcommand{\ee}{\end{equation}}
\def\ba#1\ea{\begin{align}#1\end{align}}
\newcommand{\bit}{\begin{itemize}}
\newcommand{\eit}{\end{itemize}}

\def\slashb#1{\setbox0=\hbox{$#1$}#1\hskip-\wd0\dimen0=5pt\advance
        \dimen0 by-\ht0\advance\dimen0 by\dp0\lower0.5\dimen0\hbox
          to\wd0{\hss\sl/\/\hss}}

\input{epsf}
\input{tcilatex}
\begin{document}

\title{\textbf{Short Baseline Reactor ~}$\overline{\nu }-e~$\textbf{\
Scattering Experiments and Non-Standard Neutrino Interactions at Source and
Detector}}
\author{Amir N. Khan}
\email{amir_nawaz@comsats.edu.pk}
\affiliation{Department of Physics, COMSATS IIT, Park Road, Islamabad, 44000, Pakistan}
\author{Douglas W. McKay}
\email{dmckay@ku.edu}
\affiliation{Department of Physics and Astronomy, University of Kansas, Lawrence, KS 66045}
\author{F. Tahir}
\email{farida_tahir@comsats.edu.pk}
\affiliation{Department of Physics, COMSATS IIT, Park Road, Islamabad, 44000, Pakistan }

\begin{abstract}
We investigate non-standard interaction effects in antineutrino-electron
scattering experiments with baselines short enough to ignore standard
oscillation phenomena. The setup is free of ambiguities from the
interference between new physics and oscillation effects and is sensitive to
both semileptonic new physics at the source and purely leptonic new physics
in the weak interaction scattering at the detector. We draw on the TEXONO
experiment as the model system, extending its analysis of non-standard
interaction effects at the detector to include the generally allowed
non-standard interaction phase at the detector and both non-universal and
flavor changing new physics at the reactor source. We confirm that the
current data allows for new physics constraints at the detector of the same
order as those currently published, but we find that constraints on the
source new physics are at least an order of magnitude weaker. The new
physics phase effects are at the 5\% level, noticeable in the 90\% C.L.
contour plots but not significantly affecting the conclusions. Based on
projected increase in sensitivity with an upgraded TEXONO experiment, we
estimate the improvement of sensitivity to both source and detector
non-standard interactions. We find that the bounds on source parameters
improve by an order of magnitude, but do not reach parameter space beyond
current limits. On the other hand, the detector new physics sensitivity
would push current limits by factors 5 to 10 smaller.
\end{abstract}

\date{\today}
\maketitle

\section{Introduction}

In the past several years, reactor neutrino experiments \cite{DC,DB,RENO}
and long baseline accelerator experiments \cite{T2K,MINOS} have produced
important advances in our understanding of neutrino mixing by measuring the
key mixing parameter $\theta _{13}$ by two completely independent processes.
The reactor experiments measure $\bar{\nu}_{e}$ disappearance in the flux of 
${\bar{\nu}_{e}}$s, indicating oscillation into other neutrino flavors
during the one or two kilometer trip from reactor core to detector. The
accelerator experiment measures the appearance of a $\nu _{e}$ component in
the $\nu _{\mu }$ beam from an accelerator during the hundreds of kilometers
trip from the accelerator laboratory to the detection site. Together, the
results already constrain the CP-violating phase angle in the mixing matrix 
\cite{T2K,MINOS}. Moreover, the data provide a potentially powerful probe of
non-standard interactions (NSI) \cite{nsifootnote} in the neutrino sector
involving some combination of neutrino source, propagation, and detection 
\cite{grossman, ohlsson, biggio}.

In this paper, we explore the constraints on semi-leptonic, charged current,
non-universal (NU) and flavor-changing (FC) parameters and likewise for both
NU and FC purely leptonic NSI parameters. The former appear in effective
Lagrangians for neutrino production from reactors and from accelerators and
for neutrino detection by inverse beta decay. The latter appear in neutrino
production from muon decay and from neutrino detection by $\nu e-~$or $\bar{%
\nu}e-$ scattering. We will focus on the case of \emph{very} short baseline
reactor $\bar{\nu}_{e}$ source and detection of the recoil electron from $%
\bar{\nu}_{e}+e\rightarrow \bar{\nu}_{e}+e$ scattering at the detector. We
rely heavily on the example provided by the TEXONO experiment \cite%
{texono1,texono2}, which measures the recoil electron spectrum from reactor
anti-neutrinos interacting with electrons in a CsI(Tl) detector. The
baseline is less than 30m, and the oscillation of the beam can be ignored,
thus providing an especially clean test of FC "wrong flavor" $\overline{\nu }%
_{\mu }$ or $\overline{\nu }_{\tau }$ or NU "right flavor" $\overline{\nu }%
_{e}$ from the semileptonic nuclear decays in the reactor. Baselines this
short avoid the degeneracies between NSI parameters and standard neutrino
mixing parameters that occur in analysis of data from reactor experiments
with kilometer \cite{DC,DB,RENO} or tens of kilometer baselines \cite{JUNO},
degeneracies that are touched on in several recent studies \cite{ANK, ozz,
gm}.

We extend work in Ref. \cite{texono2} by incorporating the effects of NSI
produced at the source and by including the phase dependence of the FC NSI
at the detector using the data from Texono's experiment. In Ref. \cite%
{texono2}, only the NSI at the detector in the single channel of $\bar{\nu}%
_{e}e-\ $scattering are considered. With NSI at the source, there is a
modification of the $\bar{\nu}_{e}$ component and an addition of $\bar{\nu}%
_{\mu }$ and $\bar{\nu}_{\tau }$ components, so $\bar{\nu}_{\mu }e-\ $%
scattering and $\bar{\nu}_{\tau }e-\ $scattering must be incorporated by
including NSI in the elastic, purely neutral current (NC), $\bar{\nu}_{\mu }$%
e and $\bar{\nu}_{\tau }$e cross sections, applicable for analyzing data
from any short baseline neutrino scattering experiment where the oscillation
effects are ignorable. Our "no NSI propagation effects" study complements
those that probe NSI with solar neutrino, accelerator neutrino and other
reactor neutrino experiments, which involve different combinations of NSI at
source, propagation and detection effects \cite{Davidson, Jbranco1,
Jbranco2, Forero, klos, ohlss, isodar}.

In Section II, III and IV, we define our notation, specify our cross
sections and define the flux factors that go with each cross section to
unify all the standard model (SM) plus NSI contributions to the rate at
source and detector in a single framework. Our formalism allows us to make
joint confidence level (C.L.) contours with NSI parameters at source and at
detector or at source alone and at detector alone. In Sec. IV, we apply this
formalism to the TEXONO data and check key results from Refs. \cite{texono1}
and \cite{texono2}, while in Secs. V we apply the formalism to the modeled
data based on the realistically achievable sensitivity proposed for an
upgrade of the TEXONO experiment \cite{texono3}. We recap and conclude in
Sec. VI. The Appendix A in Sec. VII briefly summarizes the reactor flux and
target density input to the recoil electron spectrum in the TEXONO
experiment. Appendix B provides a table summarizing relevant model
independent NSI parameter bounds from Ref. \cite{biggio}.

\section{Formalism of source and detector NSI and the spectrum information}

\subsection{NSI effective Lagrangians at source and detector}

In the problem we address here, the source of antineutrinos is the
semileptonic, charged-current decays of reactor nuclei. At the level of the
quark content of the nucleons, the transition d $\rightarrow $ u+e+$\bar{\nu}
$ provides the antineutrinos for the elastic $\bar{\nu}e-$ scattering
process at the detector. To allow for lepton-flavor-violating decays at the
source, we adopt the semileptonic, charged-current, effective Lagrangian 
\cite{jm1,jm2,ANK}

\begin{equation}
\mathcal{L}^{s}=-2\sqrt{2}G_{F}(\delta _{\alpha \beta }+K_{\alpha \beta })(%
\bar{l}_{\alpha }\gamma _{\lambda }P_{L}U_{\beta a}\nu _{a})(\bar{d}\gamma
^{\lambda }P_{L}u)^{\dagger }+h.c.,
\end{equation}%
where repeated flavor-basis indices "$\alpha $" and "$\beta $" and
mass-basis indices "$a$" are summed over. We confine ourselves to the
left-handed quark helicity projection case for simplicity. The inclusion of
the right handed terms adds nothing essential to our discussion. Since we
consider the neutrino-propagation baselines are only a few tens of meters
and the energies are in the MeV range, therefore oscillations play no role
and we can effectively replace $U_{\beta a}\bar{\nu}_{a}\rightarrow \bar{\nu}%
_{\beta }$ in\ making the rate calculations we present here. The complex
coefficients $K_{\alpha \beta }$ represent the relative coupling strengths
of the flavor combinations in the presence of new physics, while in the SM, $%
K_{\alpha \beta }=0.$

To represent the NSI effects in the purely leptonic sector \cite{jm1, jm2,
jm3, gb1, gb2} for the simplified elastic $\bar{\nu}e-$ scattering case of
interest, we write the effective Lagrangian as

\begin{eqnarray}
\mathcal{L}^{\ell } &=&\mathcal{L}_{NU}^{\ell }+\mathcal{L}_{FC}^{\ell } 
\notag \\
&=&-2\sqrt{2}G_{F}\sum_{\alpha }(\overline{e}\ \gamma _{\mu }\left( 
\widetilde{g}_{\alpha R}P_{R}+(\widetilde{g}_{\alpha L}+1)P_{L})e\right) (%
\bar{\nu}_{\alpha }\gamma ^{\mu }P_{L}\nu _{\alpha })  \notag \\
&&-2\sqrt{2}G_{F}\sum_{\alpha \neq \beta }\varepsilon _{\alpha \beta }^{eP}(%
\bar{e}\gamma _{\lambda }Pe)(\bar{\nu}_{\alpha }\gamma ^{\lambda }P_{L}\nu
_{\beta }).  \label{eq:lep}
\end{eqnarray}%
The first term in Eq. (2) is the NU case and the second term is the FC case.
The coefficients $\widetilde{g}_{\alpha R}$ and $\widetilde{g}_{\alpha L}$
are 
\begin{equation}
\widetilde{g}_{\alpha R}=\sin ^{2}\theta _{w}+\varepsilon _{\alpha \alpha
}^{eR}\ \emph{and}\ \ \widetilde{g}_{\alpha L}=\sin ^{2}\theta _{w}-\frac{1}{%
2}+\varepsilon _{\alpha \alpha }^{eL}.  \label{eq:defg}
\end{equation}%
Hermiticity of $\mathcal{L}^{\ell }$ requires that the NSI matrix of
parameters be Hermitian: $\epsilon _{\alpha \beta }^{eR,L}=(\epsilon _{\beta
\alpha }^{eR,L})^{\ast }$, so the FC NSI parameters are complex in general.
Adopting the commonly used "$\varepsilon $" notation for the leptonic sector
makes the distinction between source ($Ks$) and detector ($\varepsilon s$)
clear. With the effective Lagrangians defined, we are now ready to summarize
the cross sections and flux factors we need for the study of the NSI effects
at source and detector.

\subsection{$\bar{\protect\nu}_{e}-e$ differential scattering cross sections
in lab frame}

In the notation for the NSI terms defined in Eq. (2) above, the differential
cross section for the $\bar{\nu}_{e}$ - e scattering with neutrino lab
energy E$_{\nu }$ and recoil electron kinetic energy T can be summarized by
the expression

\begin{eqnarray}
\left[ \frac{d\sigma (\bar{\nu}_{e}e)}{dT}\right] _{SM+NSI} &=&\frac{%
2G_{F}^{2}m_{e}}{\pi }[\widetilde{g}_{eR}^{2}+\underset{\alpha \neq e}{%
\Sigma }|\varepsilon _{\alpha e}^{eR}|^{2}  \notag \\
&&+\left( (\widetilde{g}_{eL}+1)^{2}+\underset{\alpha \neq e}{\Sigma }%
|\varepsilon _{\alpha e}^{eL}|^{2}\right) \left( 1-\frac{T}{E_{\nu }}\right)
^{2}  \notag \\
&&-\left( \widetilde{g}_{eR}(\widetilde{g}_{eL}+1)+\underset{\alpha \neq e}{%
\Sigma }\Re \lbrack (\varepsilon _{\alpha e}^{eR})^{\ast }\varepsilon
_{\alpha e}^{eL}]\right) \frac{m_{e}T}{E_{\nu }^{2}}],
\end{eqnarray}%
which is the sum of the scattering cross sections for the three, incoherent
processes $\bar{\nu}_{e}+e\rightarrow \bar{\nu}_{e}\rightarrow e,\ \bar{\nu}%
_{e}+e\rightarrow \bar{\nu}_{\mu }+e$ and $\bar{\nu}_{e}+e\rightarrow \bar{%
\nu}_{\tau }+e$. The $\bar{\nu}_{e}+e\rightarrow \bar{\nu}_{e}+e$ cross
section is represented by the terms containing the $\widetilde{g}_{eL}$ and $%
\widetilde{g}_{eR}$ parameters. It is the coherent sum of the neutral
current and charged current contributions. The complex parameters $%
\varepsilon _{\alpha e}^{eL}$, $\alpha \neq e$ can be written either as $%
\varepsilon _{\alpha e}^{eL}=\Re \lbrack \varepsilon _{\alpha e}^{eL}]+i\Im
\lbrack \varepsilon _{\alpha e}^{eL}]$ or as $|\varepsilon _{\alpha
e}^{eL}|exp(i\phi _{\alpha e}^{eL})$, where $\phi _{\alpha e}^{eL}$ is the
phase angle of the complex quantity. Written out in more detail, the NSI
contributions are $|\varepsilon _{\alpha e}^{eR}|^{2}=(\Re \lbrack
\varepsilon _{\alpha e}^{eR}])^{2}+(\Im \lbrack \varepsilon _{\alpha
e}^{eR}])^{2}$, and similarly for $R\rightarrow L$. In the last term, $\Re
\lbrack (\varepsilon _{\alpha e}^{eR})^{\ast }\varepsilon _{\alpha
e}^{eL}]=\Re \lbrack \varepsilon _{\alpha e}^{eR}]\Re \lbrack \varepsilon
_{\alpha e}^{eL}]+\Im \lbrack \varepsilon _{\alpha e}^{eR}]\Im \lbrack
\varepsilon _{\alpha e}^{eL}]$. This notation makes it clear that when the $%
\varepsilon $ parameters are taken as real positive or negative, then the "$%
\Re $" and "$\Im $" notation can be dropped and one can drop the absolute
magnitude signs everywhere. All of the NSI studies with $\nu $ - e
scattering at the detector tacitly make this assumption \cite%
{texono2,Davidson,Jbranco1,Jbranco2,Forero,isodar}. If the parameters are
written as $|\varepsilon _{\alpha e}^{eL}|exp(i\phi _{\alpha e}^{eL})$ and $%
|\varepsilon _{\alpha e}^{eR}|exp(i\phi _{\alpha e}^{eR})$, then the
coefficient in the last term can be expressed as 
\begin{equation}
\Re \lbrack (\varepsilon _{\alpha e}^{eR})^{\ast }\varepsilon _{\alpha
e}^{eL}]=|\varepsilon _{\alpha e}^{eR}||\varepsilon _{\alpha e}^{eL}|\cos
(\phi _{\alpha e}^{eL}-\phi _{\alpha e}^{eR}).  \label{eq:phasefactor}
\end{equation}%
With this parameterization, the values of $|\varepsilon _{\alpha e}^{eR}|$
and $|\varepsilon _{\alpha e}^{eL}|$ are always positive and the sign of the
term is controlled by $\cos (\phi _{\alpha e}^{eL}-\phi _{\alpha e}^{eR})$.

To include the NSI at the reactor source, using the notation from \cite{ANK}%
, one multiplies the contribution to the rate by $|1+K_{ee}|^{2}$. Though
Ref. \cite{ANK} works only to first order in NSI parameters and drops the
highly constrained linear term $2\Re \lbrack K_{ee}]$ \cite{biggio}, in the
present calculation we must work to second order to assess the impact of the
NSI, so \emph{both} $\Im \lbrack K_{ee}]$ and $\Re \lbrack K_{ee}]$ will be
included in the non-universal (NU) case $\bar{\nu}_{e}+e\rightarrow \bar{\nu}%
_{e}+e$.

\subsection{$\bar{\protect\nu}_{\protect\mu}$ - e and $\bar{\protect\nu}_{%
\protect\tau}$ - e differential scattering cross sections in lab frame}

As just described above, the cross sections for each incoming neutrino
flavor must be multiplied by the corresponding NSI factor, $|K_{e\mu }|^{2}~ 
$for the incoming $\bar{\nu}_{\mu }$ component of the flux and similarly for
the $\bar{\nu}_{\tau }$ component. The $\bar{\nu}_{\mu }$ - e cross section
is

\begin{eqnarray}
\left[ \frac{d\sigma (\bar{\nu}_{\mu }e)}{dT}\right] _{SM+NSI} &=&\frac{%
2G_{F}^{2}m_{e}}{\pi }[\widetilde{g}_{\mu R}^{2}+\underset{\alpha \neq \mu }{%
\Sigma }|\varepsilon _{\alpha \mu }^{eR}|^{2}  \notag \\
&&+\left( \widetilde{g}_{\mu L}^{2}+\underset{\alpha \neq \mu }{\Sigma }%
|\varepsilon _{\alpha \mu }^{eL}|^{2}\right) \left( 1-\frac{T}{E_{\nu }}%
\right) ^{2}  \notag \\
&&-\left( \widetilde{g}_{\mu R~~}\widetilde{g}_{\mu L}+\underset{\alpha \neq
\mu }{\Sigma }\Re \lbrack (\varepsilon _{\alpha \mu }^{eR})^{\ast
}\varepsilon _{\alpha \mu }^{eL}]\right) \frac{m_{e}T}{E_{\nu }^{2}}].
\label{eq.4}
\end{eqnarray}

The cross section for $\bar{\nu}_{\tau }$ - e scattering is obtained by
replacing $\mu $ by $\tau $ everywhere in the above equation. The
definitions of $\widetilde{g}_{\mu R,\mu L}$ and $\widetilde{g}_{\tau R,\tau
L}$ are obvious counterparts to the definition of $\widetilde{g}_{eR,eL}$ in
Eq. (\ref{eq:defg}).

\subsection{Discussion of NSI at the source and the full NSI effects}

The distance between the source and detector in the TEXONO experiment is
less than 30m, so we will use the fact that the oscillation effects,
proportional to $\sin ^{2}(m_{i}^{2}-m_{j}^{2})L/4E_{\nu }$, are ignorable
for the range of interest, $3MeV\leq E_{\nu }\leq 8MeV$. In effect, this
means that the flavor of neutrino that is produced at the source is the same
as the flavor that reaches the detector. The factors that control the flux
of each flavor in the incoming beam produced at the source are the $%
K_{\alpha \beta }$. The TEXONO flux model is the result of a large number of
independent nuclear reactions. In the presence of NSI, the emitted flux can
be thought of as an incoherent sum of $\overline{\nu }_{e},\overline{\nu }%
_{\mu }$ and $\overline{\nu }_{\tau }$ with weights $|1+K_{ee}|^{2},|K_{e\mu
}|^{2}$ and $|K_{e\tau }|^{2}$. The source and detector NSI effects on the
rate are then expressed through the following factor, denoted by $\mathcal{F}
$, that will multiply the reactor flux and the target electron number
density to get the differential rate $\frac{dR_{X}}{dT}$, as described in
the Appendix A: 
\begin{equation}
\mathcal{F}=|1+K_{ee}|^{2}\left[ \frac{d\sigma (\bar{\nu _{e}}e)}{dT}\right]
+|K_{e\mu }|^{2}\left[ \frac{d\sigma (\bar{\nu}_{\mu }e)}{dT}\right]
+|K_{e\tau }|^{2}\left[ \frac{d\sigma (\bar{\nu}_{\tau }e)}{dT}\right] ,
\label{eq:eff}
\end{equation}%
where the cross section formulas are as given in Eqs. (4) and (6) and the
SM+NSI designation is understood.

\section{Probing model parameters with recoil electron energy spectrum data:
the TEXONO experiment}

We reproduce and recap the TEXONO experiment \cite{texono1} and it's related
analyses \cite{texono2,texono3} that are directly relevant to our NSI
parameters study. The neutrino flux spectrum and the event rate data and its
theoretical representation are briefly summarized in the Appendix A. In Ref. 
\cite{texono1}, the primary goal was an independent determination of the
weak mixing parameter $\sin ^{2}\theta _{W}$, determined strictly from low
energy, purely leptonic recoil spectrum data in the $\bar{\nu}%
_{e}+e\rightarrow \bar{\nu}_{e}+e$ elastic scattering process. The paper
stresses that this data is more sensitive to the right-handed neutral
current (NC) component in Eq. (4) than is the corresponding $\nu
_{e}+e\rightarrow \nu _{e}+e$ scattering case, where the roles of $g_{L}$
and $g_{R}$ are reversed. The $\bar{\nu}_{e}+e$ scattering is consequently
more sensitive to $g_{R}$= $\sin ^{2}\theta _{W}$. Using their flux and
binned rate spectrum \cite{Thanks}, we show the result of a $\chi ^{2}$
analysis with statistical errors only in Fig. 1(a). The 1$\sigma $ and $90\%$
C.L. lines are included for guidance. We find a best fit of $\sin ^{2}\theta
_{W}=0.251\pm 0.030$ in agreement with TEXONO's result.

\begin{figure}[tbph]
\begin{center}
\includegraphics[width=7in]{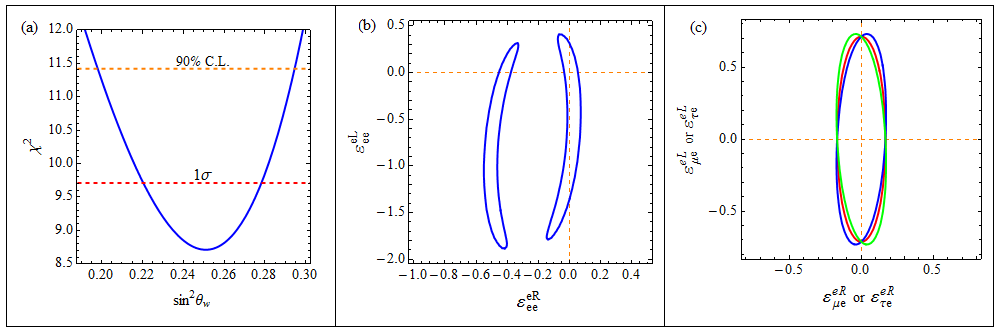}
\end{center}
\caption{{}SM $\sin ^{2}\protect\theta _{W}$ vs. $\protect\chi ^{2}$, 1(a),
and our calculation of the 90\% C. L. limits of Fig. 4(a) and 4(b) of Ref.
[13] Figs. 1(b) and 1(c). In Fig. 1(c), we show the 90\% C.L. boundary for
the fit to TEXONO rate data using Eq. (5) in the scattering cross section
Eq. (2). The blue(B), red(R) and green(G) curves, right-to-left at the top,
are for $\cos (\protect\phi _{\protect\alpha e}^{eL}-\protect\phi _{\protect%
\alpha e}^{eR})=1,\ 0$ and $-1,$ respectively. The blue curve, with $\cos (%
\protect\phi _{\protect\alpha e}^{eL}-\protect\phi _{\protect\alpha %
e}^{eR})=1,$ corresponds to that shown in Fig. 4c of Ref. [13].}
\label{texono4}
\end{figure}
Following publication of their experimental results \cite{texono1},
detailing the experiment and the results on $\sin ^{2}\theta _{W}$ and on an
upper limit of the neutrino magnetic moment, the collaboration presented
limits on NSI parameters and on couplings of unparticles to neutrinos and
electrons \cite{texono2}. Since we are pursuing an extension of the NSI
bounds to include the possibility of semi-leptonic NSI modifications to the
reactor source of $\bar{\nu}_{e}$s and the interplay with the purely
leptonic detection NSI, we are primarily interested in C.L. boundaries in
two parameter fits to the data and the joint limits obtained from these
analyses. For illustration, we check our evaluation of the 90\% C.L.
boundaries in the $\epsilon _{ee}^{eR}-\epsilon _{ee}^{eL}$ plane and,
alternatively, the $\epsilon _{e\tau }^{eR}-\epsilon _{e\tau }^{eL}$ plane,
Figs. 4a and 4b in \cite{texono2}. We show the result of this exercise in
Figs. \ref{texono4}(b) and the blue boundary, rightmost at the top, in Fig. %
\ref{texono4}(c). In both cases we find that our results and TEXONO's agree
within the ability to read off values along the contours. We show the 90\%
C.L. projections of these plots on the individual axes for the two cases in
Table \ref{Table1}. The red and green curves, center and leftmost in Fig. %
\ref{texono4}(c) are examples of other phase choices, as we discuss in
subsection IIIA. For the NU case of $\epsilon _{ee}^{eR}-\epsilon _{ee}^{eL}$
plane, we quote the right-hand solution values, since both the R and L
limits are the most stringent for this solution. The FC case assumes the NSI
parameters are purely real. There is no degeneracy in this case, and the
projected individual two parameter limits are straightforward. The weak
correlation between the R and L NSI parameters is due to the small R-L NSI
interference term. Though our contour agrees with \cite{texono2} and our $%
\epsilon _{\tau e}^{eR}$ bounds agree with the ones quoted in their Table I,
our limits on $\epsilon _{\tau e}^{eL}$ are somewhat smaller.

\begin{table*}[tbph]
\begin{center}
\begin{tabular}{l|l|l}
\hline\hline
Figure No. & R-Parameter Bounds & L-Parameter Bounds \\ \hline
$1(b)$ & $-0.15<\varepsilon _{ee}^{eR}<0.08$ & $-1.79<\varepsilon
_{ee}^{eL}<0.41$ \\ 
$1(c)$ & $-0.18<\varepsilon _{\alpha e}^{eR}<0.18$ & $-0.76<\varepsilon
_{\alpha e}^{eL}<0.76$ \\ \hline\hline
\end{tabular}%
\end{center}
\caption{{}Bounds at $90\%\ $C.L. obtained from Fig. \protect\ref{texono4}%
(b) and \protect\ref{texono4}(c) in the absence of any source NSI where $%
\protect\alpha =\protect\mu ,~\protect\tau .$ }
\label{Table1}
\end{table*}

\subsection{Role of the detector NSI phases in determining the C.L.
boundaries}

The R-L interference term in the differential cross sections depends on the
FC NSI parameter phases, as displayed for the case $\bar{\nu}%
_{e}+e\rightarrow \bar{\nu}+e$ in Eq. (\ref{eq:phasefactor}). From the point
of view of this general formula, the blue boundary, rightmost at the top, in
Fig. \ref{texono4}(c) can be interpreted as the composite of the cases $\phi
_{\mu e}^{eR}=\phi _{\mu e}^{eL}=0$, where $\epsilon _{\mu e}^{eR}$ and $%
\epsilon _{\mu e}^{eL}$ are both real and positive, $\phi _{\mu e}^{eR}=\pi $
and $\phi _{\mu e}^{eL}=0$, where $\epsilon _{\mu e}^{eR}$ is real and
negative and $\epsilon _{\mu e}^{eL}$ is real and positive, $\phi _{\mu
e}^{eR}=\phi _{\mu e}^{eL}=\pi $, where $\epsilon _{\mu e}^{eR}$ and $%
\epsilon _{\mu e}^{eL}$ are both real and negative, and, finally, $\phi
_{\mu e}^{eR}=0$ and $\phi _{\mu e}^{eL}=\pi $, where $\epsilon _{\mu
e}^{eR} $ is real and positive and $\epsilon _{\mu e}^{eL}$ is real and
negative. Alternatively, it can be interpreted as the composite of cases
where 0 and ${\pi }$ are replaced with $\pi /2$ and $3\pi /2$ and real
replaced with imaginary. Because the R-L interference term is suppressed by
the factor $m_{e}T/E_{\nu }^{2}$ and $E_{\nu }\geq $ 3 MeV, the changes in
the parameter boundaries as the phase differences range from 0 to $\pi $ are
small, as shown in Fig. \ref{texono4}(c). Conclusions about allowed
boundaries for NSI parameters for the range of energies of interest in
reactor experiments are affected very little in this analysis, but for
experiments with significantly lower energy radioactive sources or for low
energy solar neutrino experiments such as Borexino \cite{borex}, the R-L
correlation term can be relatively larger and the phase effects may be
important. For present purposes, we illustrate the range of effects that
change of phases can make on the C.L. boundary in Fig. \ref{texono4}(c). The
small changes in boundaries is shown in the figure by the difference between
the blue, red and green curves, corresponding to $\cos (\phi _{\alpha
e}^{eL}-\phi _{\alpha e}^{eR})$. = 1, 0 and -1, reading from right to left
at the top of the figure. As one sees, the correlation disappears for the
case that $\cos (\phi _{\alpha e}^{eL}-\phi _{\alpha e}^{eR})$ =0, the red,
middle curve. The R-L correlation term vanishes in this case because the R
and L parameters are $\pi /2$ out of phase; one can be real and the other
imaginary, for example.

\section{Interplay between $K_{\protect\alpha \protect\beta }$ (source)$\ $%
and\ $\protect\epsilon _{\protect\alpha \protect\beta }^{eR,L}$ (detector)
NSI parameters}

In this section we take pairs of source and detector NSI coefficients to
survey the 90\% C.L. boundaries in the various two-parameter spaces. We
focus on the bounds on the source parameters and assess the strength of the
bounds found to the bounds currently available in the literature. At the
same time, we check for consistency of the bounds on the detector NSI
parameters with those found by TEXONO \cite{texono2}, which we checked in
the preceding section.

\begin{figure}[tbph]
\begin{center}
\includegraphics[width=7in]{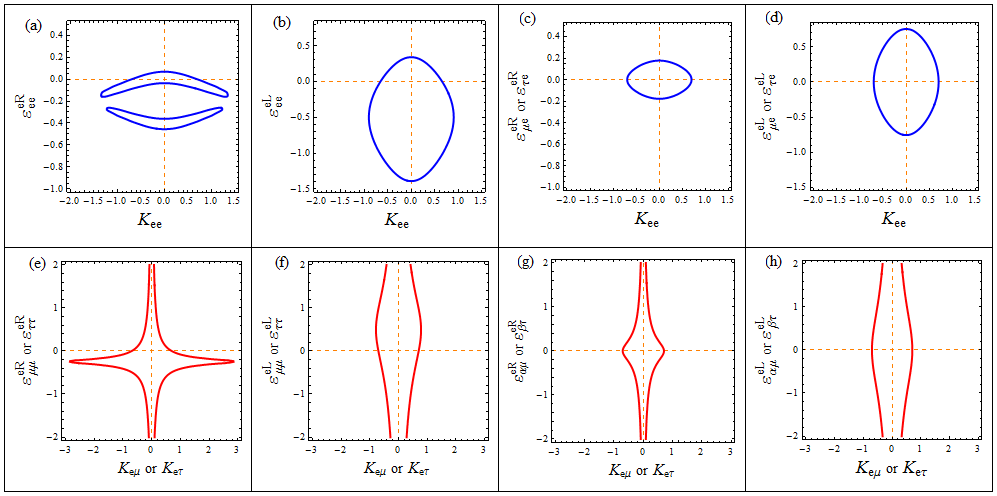}
\end{center}
\caption{{}C.L boundary regions for TEXONO\ \emph{realistic} data. \textbf{%
Upper Panel:} Correlation between the source NSI parameter $(K_{ee})$ and
the corresponding detector NSI parameters ($\protect\varepsilon _{ee}^{R,\ L}
$ and $\protect\varepsilon _{\protect\alpha e}^{R,\ L},$ where $\protect%
\alpha \ $=\ $\protect\mu $, $\protect\tau $) at 90\% C.L.. See the text for
details. \textbf{Lower Penal}: Correlation between the source NSI parameter $%
(K_{e\protect\alpha })$ and the corresponding detector NSI parameters ($%
\protect\varepsilon _{\protect\mu \protect\mu }^{R,\ L}$ ,\ $\protect%
\varepsilon _{\protect\tau \protect\tau }^{R,\ L}$and $\protect\varepsilon _{%
\protect\alpha \protect\mu }^{R,\ L},\ \protect\varepsilon _{\protect\beta 
\protect\tau }^{R,\ L}$ where $\protect\alpha \ $=\ $e$, $\protect\tau $ and$%
~\protect\beta =$ $e,\protect\mu ~$) at 90\% C.L.. See the text for details.}
\label{Keps1}
\end{figure}

Since the current bounds on the real part of $K_{ee}$ are of the order $%
10^{-3}$, as given in \cite{biggio} and found independently from Daya Bay
data by \cite{gm}, and these are much tighter than we can imagine providing
with the current analysis based on the TEXONO data, we assume $K_{ee}$ is
purely imaginary in this section. Consequently the source coefficient in the
case of incident $\bar{\nu}_{e}$ in Eq. (\ref{eq:eff}) is $K_{ee}^{2}=1+(\Im
\lbrack K_{ee}])^{2}$. Bounds found will then refer to $\Im \lbrack K_{ee}]$%
. Fig. \ref{Keps1} shows the 90\% C.L. boundaries for the fits to the TEXONO
data as parameterized by one source NSI coefficient and one detector
coefficient with all of the other NSI coefficients set to zero. 
From the 90\% C.L. contours shown in Fig. \ref{Keps1}, we can determine the
90\% C.L. bounds on the source NU $K_{ee}$ parameter and any of the $%
\epsilon _{\alpha e}^{eR,L}$ at the detector by projecting onto the
parameter axes for each contour. We find the limits on the NU parameters
quoted in Table \ref{tabKeps1}. In all of the cases involving the source FC
semileptonic NSI parameters $K_{\alpha \beta }$, there is no bound on any of
the leptonic, detector NSI parameters $\epsilon _{\alpha \beta }^{eR,L}$ as $%
K_{\alpha \beta }\rightarrow 0$, because the source is receiving only $\bar{%
\nu}_{e}$ flux in this limit. In this sense, the parameters $K_{\alpha \beta
}$ and $\epsilon _{\alpha \beta }^{eR,L}$ are highly correlated. There is
still the possibility for placing upper bounds on the $K_{e\alpha }$
parameters in this case if the detector NSI parameters are constrained to be
smaller than their current bounds \cite{biggio}, which are near zero on the
scale of Fig. \ref{Keps1}. We can then place upper 90\% C.L. bounds on the
values of $K_{e\mu }$ or $K_{e\tau }$ for the special case where detector
NSI $\epsilon _{\mu \mu }^{eR,\ L}=\epsilon _{\alpha \mu }^{eR,\ L}$ $=0$,
and likewise for $\mu \rightarrow \tau $. These one-parameter-at-a time
upper bounds on $K_{e\mu }$ or $K_{e\tau }$, the type commonly reported in
the literature, are the bounds we quote in Table \ref{tabKeps1} \cite%
{nocorrel}.

\begin{table*}[tbph]
\begin{center}
\begin{tabular}{l|l|l}
\hline\hline
Figure No. & NSI Parameters at Source & \ NSI\ Parameters at Detector \\ 
\hline
$2(a)$ & $-1.35<\func{Im}K_{ee}\ <1.35$ & $-0.17<\varepsilon _{ee}^{eR}<0.07$
\\ 
$2(b)$ & $-0.9\ \ <\func{Im}K_{ee}~<0.9$ & $-1.4\ \ <\varepsilon _{ee}^{eL}\
<0.34$ \\ 
$2(c)$ & $-0.72<\func{Im}K_{ee}\ <0.72\ $ & $-0.18<\varepsilon _{\alpha
e}^{eR}<0.18$ \\ 
$2(d)$ & $-0.72<\func{Im}K_{ee}\ <0.72$ & $-0.76<\varepsilon _{\alpha
e}^{eL}<0.76$ \\ 
$2(e)$ & $-0.72<\func{Im}K_{e\mu }\ $or$\ \func{Im}K_{e\tau }\ <0.72\ $ & $%
\varepsilon _{\mu \mu }^{eR}\ $and $\varepsilon _{\tau \tau }^{eR}~$are
unbounded \\ 
$2(f)$ & $-0.72<\func{Im}K_{e\mu }\ $or $\func{Im}K_{e\tau }\ <0.72$ & $%
\varepsilon _{\mu \mu }^{eL}\ $and $\varepsilon _{\tau \tau }^{eL}~$are
unbounded \\ 
$2(g)$ & $-0.72<\func{Im}K_{e\mu }\ $or$\ \func{Im}K_{e\tau }\ <0.72$ & $%
\varepsilon _{\alpha \mu }^{eR}\ $and $\varepsilon _{\beta \tau }^{eR}~$are
unbounded \\ 
$2(h)$ & $-0.72<\func{Im}K_{e\mu }\ $or$\ \func{Im}K_{e\tau }\ <0.72$ & $%
\varepsilon _{\alpha \mu }^{eL}\ $and $\varepsilon _{\beta \tau }^{eL}~$are
unbounded \\ \hline\hline
\end{tabular}%
\end{center}
\caption{Bounds at $90\%\ $C.L. obtained from Fig. \protect\ref{Keps1} \
where $\protect\alpha =e,~\protect\tau $ and $\protect\beta =e,~\protect\mu %
. $}
\label{tabKeps1}
\end{table*}

Briefly summarized, the results of this study based on the published TEXONO
data show that the sensitivity to reactor \emph{source} NSI, $K_{\alpha
\beta }$, is at least an order of magnitude less than the sensitivity of the
data used to establish the currently available bounds. On the other hand,
the sensitivity to \emph{detector} NSI is of the same order of magnitude as
the current bounds for the right handed NSI couplings, though much less for
the left-handed couplings. The future improvements in sensitivity, as
envisioned by the TEXONO collaboration \cite{texono2,texono3}, should change
this situation considerably, and we turn to this consideration in the next
section.

\section{Future Prospects}

\begin{figure}[tbph]
\begin{center}
\includegraphics[width=7in]{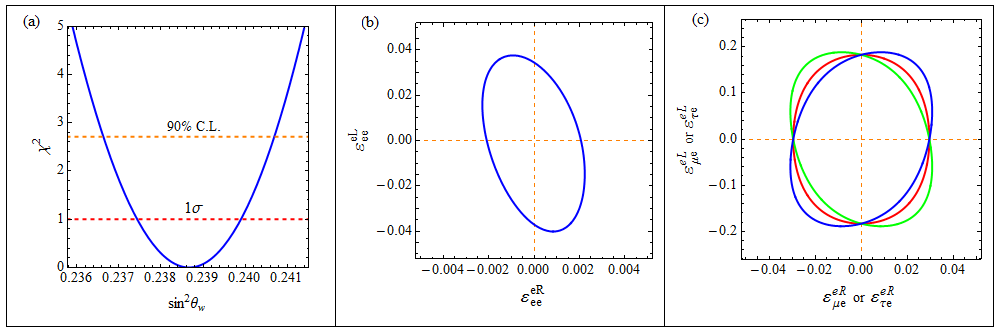}
\end{center}
\caption{{}The $90\%$ C.L. contours for NU and FC target NSI for the model
data described in the text. The source NSI are set to zero.}
\label{Figureimproved20}
\end{figure}

In this section we study the future prospects for tightening the source and
detector NSI parameter bounds by adopting the projected "realistic and
feasible" improvements in statistical sensitivities reported in Table 2 and
the related text in Ref. \cite{texono3}. Their essential point is that
statistical uncertainty of the measured value of $\sin ^{2}\theta _{W}$ can
realistically be reduced to $\pm 0.0013$. We follow the experimental setup
from Ref. \cite{texono1, texono2} and generate our data in 10 energy bins,
each of step 0.5 Mev . We generate our "data model" by assuming that the
best fit value turns out to be $\sin ^{2}\theta _{W}=0.2387$ \cite{Jerler},
the value cited for comparison to their experimental fit value $\sin
^{2}\theta _{W}=0.251$ by Ref. \cite{texono1}. We define our model $\chi
^{2} $ distribution by forming

\begin{equation}
\chi ^{2}=\underset{i}{\tsum }\left( \frac{R_{NSI}-R_{SM}}{\Delta _{stat}}%
\right) _{i}^{2}
\end{equation}%
where $R_{SM}\ $is the data model rate, R$_{NSI}$ is predicted event rate
with all unknown NSI parameters and$\ \Delta _{stat}$ is the statistical
uncertainty over each bin. We define $\ \Delta _{stat}$ as the deviation
from the central value $R_{SM}\ $within 1$\sigma \ $statistical uncertainty,
obtained by evaluating the rate with $\sin ^{2}\theta _{W}=0.2387.$ To
achieve a fit to the 10 bins of rate data that yields the projected
uncertainty of $\pm 0.0013$ for $\sin ^{2}\theta _{W},$ we find that
evaluating the rates in each bin with $\sin ^{2}\theta _{W}$ \ roughly $(%
\sqrt{10}\simeq 3)\times \pm 0.0013$ per bin and taking the average
deviation from the central value yields a data set whose uncertainties are
consistent with expectations \cite{texono3}. We take this model set as the
basis for estimated future sensitivity to NSI \cite{isodar}. The results
shown in Fig. \ref{texono4} are redone using future prospects data in Fig. %
\ref{Figureimproved20} and the bounds obtained at 90\% C.L. are given in
Table \ref{tableimp20} \cite{isodar}.

\begin{table*}[tbph]
\begin{center}
\begin{tabular}{l|l|l}
\hline\hline
Figure No. & \ \ R-Parameter Bounds & \ \ \ \ \ \ \ L-Parameter Bounds \\ 
\hline
$3(b)$ & $-0.0023<\varepsilon _{ee}^{eR}<0.0023$ & $-0.04<\varepsilon
_{ee}^{eL}<0.04$ \\ 
$3(c)$ & $\ \ -0.03<\varepsilon _{\alpha e}^{eR}<0.03$ & $-0.19<\varepsilon
_{\alpha e}^{eL}<0.19$ \\ \hline\hline
\end{tabular}%
\end{center}
\caption{{}Bounds at $90\%\ $C.L. obtained from Fig. \protect\ref%
{Figureimproved20}(b) and \protect\ref{Figureimproved20}(c) in the absence
of any source NSI where $\protect\alpha =\protect\mu ,~\protect\tau .$ }
\label{tableimp20}
\end{table*}

From Fig. \ref{Figureimproved20} and the bounds summarized in Table \ref%
{tableimp20}, we see immediately the impact of improved sensitivity to the
presence of NSI at the detector in the removal of the degeneracy in the $%
\epsilon _{ee}^{eR}$ vs. $\epsilon _{ee}^{eL}$ plot when compared to Fig. %
\ref{texono4}. The purely leptonic NU and FC new physics effects can be
probed with up to two orders of magnitude higher refinement in the right
handed lepton sector and up to an order of magnitude more refinement in the
left-handed sector. With comparable experimental sensitivity in a $\nu
_{e}+e\rightarrow \nu _{e}+e$ experiment, a complimentary result with the 
\emph{left-handed} sector being favored could be achieved \cite{LENA,
garcesetal} .

\begin{figure}[tbph]
\begin{center}
\includegraphics[width=7in]{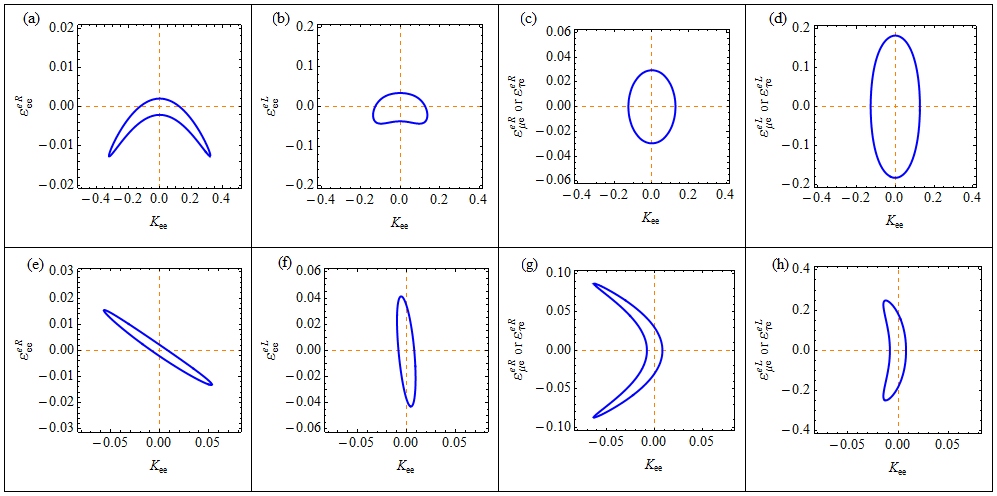}
\end{center}
\caption{{}C.L boundary regions for \emph{future prospects} data. \textbf{%
Upper Panel:} Correlation between the source NSI parameter $(K_{ee})$ and
the corresponding detector NSI parameters ($\protect\varepsilon _{ee}^{R,\ L}
$ and $\protect\varepsilon _{\protect\alpha e}^{R,\ L},$ where $\protect%
\alpha \ $=\ $\protect\mu $, $\protect\tau $) at 90\% C.L.. See the text for
details. \textbf{Lower Panal}: Correlation between the source NSI parameter $%
(K_{ee})$ and the corresponding detector NSI parameters ($\protect%
\varepsilon _{\protect\mu \protect\mu }^{R,\ L}$ ,\ $\protect\varepsilon _{%
\protect\tau \protect\tau }^{R,\ L}$and $\protect\varepsilon _{\protect%
\alpha \protect\mu }^{R,\ L},\ \protect\varepsilon _{\protect\beta \protect%
\tau }^{R,\ L}$ where $\protect\alpha \ $=\ $e$, $\protect\tau $ and$~%
\protect\beta =$ $e,\protect\mu ~$) at 90\% C.L.. See the text for details.}
\label{Figureimproved1}
\end{figure}

\begin{table*}[tbph]
\begin{center}
\begin{tabular}{l|l|l}
\hline\hline
Figure No. & NSI Parameters at Source & \ NSI\ Parameters at Detector \\ 
\hline
$4(a)$ & $-0.33<\func{Im}K_{ee}<0.33$ & $-0.013<\varepsilon _{ee}^{eR}<0.002$
\\ 
$4(b)$ & $-0.14<\func{Im}K_{ee}<0.14$ & $-0.045<\varepsilon _{ee}^{eL}<0.036$
\\ 
$4(c)$ & $-0.13<\func{Im}K_{ee}<0.13$ & $-0.03\ \ <\varepsilon _{\alpha
e}^{eR}<0.03$ \\ 
$4(d)$ & $-0.13<\func{Im}K_{ee}<0.13$ & $-0.18\ \ <\varepsilon _{\alpha
e}^{eL}<0.18$ \\ 
$4(e)$ & $-0.057<\func{Re}K_{ee}<0.054$ & $-0.013<\varepsilon
_{ee}^{eR}<0.016$ \\ 
$4(f)$ & $-0.01\ \ <\func{Re}K_{ee}<0.01$ & $-0.043<\varepsilon
_{ee}^{eL}<0.042$ \\ 
$4(g)$ & $-0.064<\func{Re}K_{ee}<0.007$ & $-0.086<\varepsilon _{\alpha
e}^{eR}<0.086$ \\ 
$4(h)$ & $-0.015<\func{Re}K_{ee}<0.008$ & $-0.25\ \ <\varepsilon _{\alpha
e}^{eL}<0.25$ \\ \hline\hline
\end{tabular}%
\end{center}
\caption{{}Bounds at 90\% C. L. obtained from Fig. (\protect\ref%
{Figureimproved1}) where $\protect\alpha =e,~\protect\tau $ and $\protect%
\beta =e,~\protect\mu .$}
\label{tableimp1}
\end{table*}

Turning to the cases where the NSI can be active at \emph{both} the source
and detector, we study the parameter spaces of combined source-detector
pairs in Fig. \ref{Figureimproved1} and in accompanying Table \ref{tableimp1}%
. The sensitivity to the combinations improves typicallyby factors of 5 to
10 in both source and detector probes compared to the bound shown in Fig. %
\ref{Keps1} and Table \ref{tabKeps1}. Comparing to current bounds in our
Appendix B, for example, we find that the bound \ on$\ \varepsilon
_{ee}^{eR}\ \ $in entry 4(e) is a factor 10 below the bound given there,
while the bound on $\varepsilon _{\tau e}^{eR}$ given in entry 4(g) is a
factor of 5 below its bound quoted in \cite{biggio}. In the case of NU $%
K_{ee}$ couplings, the constraints are becoming competitive with those
published \cite{biggio}, being within about a factor of 3 for both the
imaginary part (top 4 rows) and the real part (bottom 4 rows) of $K_{ee}$.
Looking at entry 4(c) or 4(d) in Table \ref{tableimp1}, we find that $|\func{%
Im}K_{ee}|\ <0.13$ , compared to the current best bound of $0.041$, which is
also the best bound for $|\func{Im}K_{e\tau }|\ $, compared \ to our bound
of $0.1$ shown in Table \ref{tableimpFullForm}. Thus, an upgraded TEXONO
experiment could provide independent confirmation of the bounds on these
parameters, but would not probe new parameter space in the search for new
physics. Similarly, as shown in Fig. 5 and Table V, the bounds on the FC
semileptonic parameters $K_{e\mu }~$and $K_{e\tau }\ $achievable by an
upgraded TEXONO experiment are within a factor 2 or 3 of the current bounds
and possibly provide independent support, but not reach new regions in their
parameter space.

Though the FC $K_{e\alpha }$ vs. $\epsilon _{\alpha \mu }$ or $\epsilon
_{\alpha \tau }$ studies, Fig. 5, provide no bounds on the $\epsilon $s
because the "wrong flavor" source neutrinos are zero in the $K_{e\mu }$ or $%
K_{e\tau }\rightarrow 0$ limit, the $|\epsilon _{\alpha e}^{eR,L}|$ limits
in Table IV apply to $|\epsilon _{e\alpha }^{eR,L}|$ because of the
Hermiticity constraint $\epsilon _{\alpha \beta }^{eR,L}=\epsilon _{\beta
\alpha }^{eR,L}$, as noted after Eq. (3).

\begin{figure}[tbph]
\begin{center}
\includegraphics[width=7in]{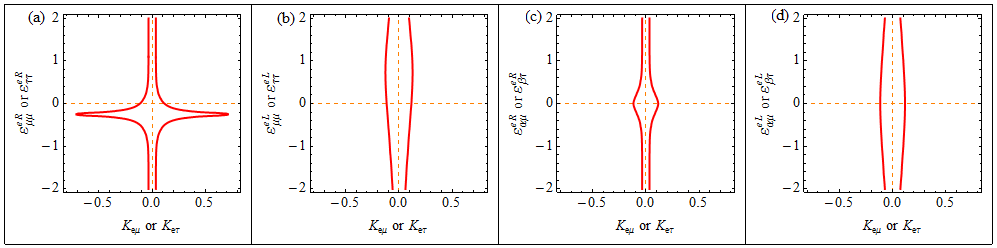}
\end{center}
\caption{ Correlation between the source NSI parameter $(\func{Re}K_{e%
\protect\mu }\ $and$\ \func{Re}K_{e\protect\tau })$ and the corresponding
detector NSI parameters ($\protect\varepsilon _{\protect\mu \protect\mu %
}^{R,\ L}$ ,\ $\protect\varepsilon _{\protect\tau \protect\tau }^{R,\ L}$and 
$\protect\varepsilon _{\protect\alpha \protect\mu }^{R,\ L},\ \protect%
\varepsilon _{\protect\beta \protect\tau }^{R,\ L}$ where $\protect\alpha \ $%
=\ $e$, $\protect\tau $ and$~\protect\beta =$ $e,\protect\mu ~$) at 90\%
C.L.. See the text for details.}
\label{FigureimprovedFullForm}
\end{figure}

\begin{table*}[tbph]
\begin{center}
\begin{tabular}{l|l|l}
\hline\hline
Figure No. & NSI Parameters at Source & \ NSI\ Parameters at Detector \\ 
\hline
$5(a)$ & $-0.1<K_{e\mu }\ $or$~K_{e\tau }<0.1$ & $\varepsilon _{\mu \mu
}^{eR}~$and $\varepsilon _{\tau \tau }^{eR}~$are unbounded \\ 
$5(b)$ & $-0.1<K_{e\mu }\ $or$~K_{e\tau }<0.1$ & $\varepsilon _{\mu \mu
}^{eL}~$and $\varepsilon _{\tau \tau }^{eL}~$are unbounded \\ 
$5(c)$ & $-0.1<K_{e\mu }\ $or$~K_{e\tau }<0.1$ & $\varepsilon _{\alpha \mu
}^{eR}~$and $\varepsilon _{\beta \tau }^{eR}~$are unbounded \\ 
$5(d)$ & $-0.1<K_{e\mu }\ $or$~K_{e\tau }<0.1$ & $\varepsilon _{\alpha \mu
}^{eL}~$and $\varepsilon _{\beta \tau }^{eL}~$are unbounded \\ \hline\hline
\end{tabular}
\ \ 
\end{center}
\caption{{}Bounds obtained from Fig. (\protect\ref{FigureimprovedFullForm})
at 90\% C. L. where $\protect\alpha =e,~\protect\tau $ and $\protect\beta %
=e,~\protect\mu .$ All the source NSI parameters $K_{\protect\alpha \protect%
\beta }$ are either pure real or imaginary.}
\label{tableimpFullForm}
\end{table*}

\section{Summary and Conclusions}

We have explored the consequences of adding new physics effects at the
reactor source in a $\bar{\nu}+e\rightarrow \bar{\nu}+e$ scattering
experiment. We have used the data from the TEXONO experiment and also a
model data based on their projected improved sensitivity in a future
upgrade. This experiment has the virtue that its 30m baseline does not allow
for oscillation effects at the detector, so that any new physics at the
source is not degenerate with oscillation effects during propagation. After
developing the needed framework in Secs. II and III, where we explicitly
include the NSI phases in the FC leptonic, detector parameters, we reviewed
the 90\% C.L. boundaries presented in Ref. \cite{texono2} in Sec IV, but
including the phase effects on the boundary in the FC, $\epsilon _{e\mu
}^{eR}-\epsilon _{e\mu }^{eL}$ parameter space. We checked that we properly
reproduce the boundaries and the value, and statistical error of the TEXONO
examples, but adding the small but noticeable dependence on the choice of
phases for the FC detector NSI parameters, filling a gap in the literature.
The effects on the bounds one derives is at the 5\% level. In lower energy
experiments with sufficient statistics, this phase effect may be more
striking as the coefficient of the correlation term becomes larger relative
to the other terms contributing to the rate.

Including the NSI at the reactor source, we surveyed examples of the
interplay between the source and detector effects with a series of source
vs. detector 90\% C.L. boundaries based on the TEXONO data. We find that the
right (R) parameter bounds on the detector NSI parameters $\epsilon _{\alpha
e}^{eR}$, $\alpha =e$, $\mu $ and $\tau $, are about the same as the current
best bounds, as summarized from Ref. \cite{biggio} in our Appendix B, but
the corresponding left (L) parameter bounds are factors 5 - 10 larger. All
of the bounds on the source, $K_{\alpha \beta }$ parameters are one-to-two
orders of magnitude larger than best current bounds. Because the source FC
parameters must be non-zero for a bound on the detector parameters $\epsilon
_{e\alpha }^{eL,R}$ to exist, no meaningful bounds can be placed
independently on the latter, but they differ only by a phase from the $%
\epsilon _{\alpha e}^{eL,R}$ parameters, as noted after Eq. (\ref{eq:defg}),
so the bounds on detector parameters listed in rows 2(c), and 2(d) in Table %
\ref{Keps1} apply as well to the detector parameters in rows 2(g) and 2(h)
when $\alpha $ and $\beta $ = e.

Turning to the companion study of our model data based on the estimated
future improvements in an upgraded TEXONO experiment, we basically repeated
the exercises of Secs. IV and V to survey the parameter spaces in
anticipation of this upgrade. Compared to the bounds based on current data
our estimates of future, high sensitivity data show that an order of
magnitude increase in the level of sensitivity to source and detector NSI
parameters is achievable compared to the sensitivity with the current TEXONO
data. This brings the bounds on detector NSI parameters well below current
bounds in all but the case of $\epsilon _{e\mu }^{eL}$, which is the same as
the current bound. Our new approach to bounding the charged current,
semileptonic NSI at the source results in projected bounds that are
comparable to the current ones. The very feature that makes this class of
ultra-short-baseline experiments especially clean for probing the source NSI
parameters, namely the lack of interference with neutrino mixing amplitudes,
makes it less sensitive. The parameters of interest appear as the modulus 
\emph{squared} in the FC case, while in the NU case, the interference with
the SM contribution gives a boost to the sensitivity to the real part of $%
K_{ee}$, which has a very tight bound already, coming from CKM unitarity or
lepton universality and for the same reason.

To conclude, we see that the currently envisaged upgrade to the TEXONO
experiment promises to probe an order of magnitude deeper into the
right-handed, leptonic NSI parameter space. To improve the sensitivity to
the left-handed, leptonic NSI couplings, high intensity, short baseline $\nu
_{e}$ experiments with large targets, along the lines of the LENA project 
\cite{LENA, garcesetal} will be needed. To delve deeper into the
semileptonic, charged current parameter space with a reactor, antineutrino
source, a third generation of the TEXONO type of experiment would be needed,
since we find that the current plans would only bring bounds to the level of
those currently available. Otherwise, oscillation experiments with
interference between the relevant NSI parameters and oscillation amplitudes
involving standard oscillation parameters, independently measured and known
to high accuracy would be needed, as remarked in \cite{ANK}.

\section{Appendix}

\subsection{Reactor neutrino spectrum and event rate: the TEXONO Experiment
at Kuo-Sheng}

\begin{figure}[tbh]
\begin{center}
\includegraphics[width=3in]{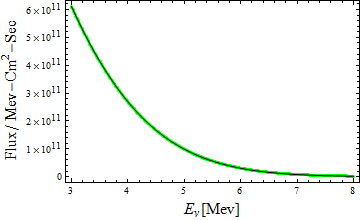}
\end{center}
\caption{Typical antineutrino spectrum at 28m from core at Kuo-Sheng. The
green curve is the data and the black curve inside it is the fit.}
\label{fig:nuspectrum}
\end{figure}

The reactor antineutrinos spectrum produced at Kuo-Sheng Nuclear Reactor is
given in Fig. \ref{fig:nuspectrum}. 
We find the following fit function for the reactor neutrino spectrum between
3Mev to 8Mev,

\begin{equation}
\frac{d\ \phi (\overline{\nu }_{e})}{dE_{\nu }}=\sum_{0}^{6}\frac{a_{n}}{%
(E_{\nu })^{n}}  \label{eqn4}
\end{equation}%
where the fit parameters $a_{0},a_{1}...a_{6}\ $have the values given in
Table \ref{Tablspectfit},

\begin{table*}[tbph]
\begin{center}
\begin{tabular}{lllllll}
\hline\hline
$\ \ \ \ \ \ \ \ \ \ \ \ a_{0}$ & \ \ \ \ \ \ \ \ \ \ \ $a_{1}$ & $\ \ \ \ \
\ \ \ \ \ \ a_{2}$ & $\ \ \ \ \ \ \ \ \ \ \ a_{3}$ & $\ \ \ \ \ \ \ \ \ \
a_{4}$ & $\ \ \ \ \ \ \ \ \ \ \ a_{5}$ & $\ \ \ \ \ \ \ \ \ \ a_{6}$ \\ 
\hline
$-1.23779\ 10^{12}$ & $\ \ \ \ 3.72889\ 10^{13}$ & $-4.38337\ 10^{14}$ & $\
2.52571\ 10^{15}$ & $-7.4559\ \ 10^{15}$ & $1.11498\ 10^{16}$ & $-6.74817\
10^{15}$ \\ \hline\hline
\end{tabular}%
\end{center}
\caption{{}Fit parameters for the neutrino spectrum.}
\label{Tablspectfit}
\end{table*}

The experimentally observed event rate (R$_{E}$) is then compared with the
theoretically modeled or expected event rate (R$_{X}$). The differential
rate with respect to T, kinetic energy of the recoil electron, is%
\begin{equation}
\frac{dR_{X}}{dT}=\rho _{e}\int_{T}^{E_{\nu }^{max}}\mathcal{F}(E_{\nu })\ 
\frac{d\ \phi (E_{\nu })}{dE_{\nu }}\ dE_{\nu },  \label{eq:difrate}
\end{equation}%
so the rate integrated over the $i^{th}$ bin in T is 
\begin{equation}
R_{X}^{i}=\int_{T(i)}^{T(i+1)}\frac{dR_{X}}{dT}.  \label{eq:binrate}
\end{equation}%
Here $\rho _{e}$\ is the electron number density per kg of target mass of
CsI(TI), and $\frac{d\phi (E_{\nu })}{dE_{\nu }}\ $is the neutrino spectrum
as given in Eq. (\ref{eqn4}) and $\mathcal{F}(E_{\nu })$ is the factor
containing the NSI detector cross sections and the corresponding NSI source
parameter coefficients, as given in Eq. (\ref{eq:eff}).

We use the following definition of $\chi ^{2}$ from ref. \cite{texono2} to
perform the minimum-$\chi ^{2}$ fit, 
\begin{equation}
\chi ^{2}={\sum_{i}}\left( \frac{R_{E}^{i}-R_{X}^{i}}{\Delta _{stat}^{i}}%
\right) ^{2},
\end{equation}%
where $R_{E}^{i}\ $and\ $R_{X}^{i}$ are the experimental and expected event
rates over the $i^{th}$ data bin and $\Delta _{stat}^{i}$ is the
corresponding statistical uncertainty of the measurement.

\subsection{Bounds of Ref. \protect\cite{biggio}}

\begin{table*}[tbph]
\begin{center}
\begin{tabular}{l|l}
\hline\hline
NSI Parameters at Source & \ NSI\ Parameters at Detector \\ \hline
$|K_{ee}|<0.041$ & $|\varepsilon _{ee}^{eR}|<0.14,\ \ \ \ \ |\varepsilon
_{ee}^{eL}|<0.06$ \\ 
$-$ & $|\varepsilon _{e\mu }^{eR}|<0.10,\ \ \ \ \ |\varepsilon _{e\mu
}^{eL}|<0.10$ \\ 
$-$ & $|\varepsilon _{e\tau }^{eR}|<0.27,\ \ \ \ \ |\varepsilon _{e\tau
}^{eL}|<0.4$ \\ 
$|K_{e\mu }|<0.025\ $ & $|\varepsilon _{\mu e}^{eR}|<0.10,\ \ \ \ \
|\varepsilon _{\mu e}^{eL}|<0.10$ \\ 
$-$ & $|\varepsilon _{\mu \mu }^{eR}|<0.03,\ \ \ \ \ |\varepsilon _{\mu \mu
}^{eL}|<0.03$ \\ 
$-$ & $|\varepsilon _{\mu \tau }^{eR}|<0.10,\ \ \ \ \ |\varepsilon _{\mu
\tau }^{eL}|<0.10$ \\ 
$|K_{e\tau }|<0.041$ & $|\varepsilon _{\tau e}^{eR}|<0.27,\ \ \ \ \
|\varepsilon _{\tau e}^{eL}|<0.4$ \\ 
$-$ & $|\varepsilon _{\tau \mu }^{eR}|<0.10,\ \ \ \ \ |\varepsilon _{\tau
\mu }^{eL}|<0.10$ \\ 
$-$ & $|\varepsilon _{\tau \tau }^{eR}|<0.4,\ \ \ \ \ \ \ |\varepsilon
_{\tau \tau }^{eL}|<0.16$ \\ \hline\hline
\end{tabular}%
\end{center}
\caption{{}Bounds at 90\% C.L. taken from Eq. (44) and Eq. (45)\ of \ Ref. 
\protect\cite{biggio} for comparison. Notice that we have used our notation
for their bounds for convenience. It should be noted that there is a
separate upper bound $\func{Re}K_{ee}\sim 10^{-3}$ from the CKM unitarity
and lepton universality constraints.}
\label{biggiotab}
\end{table*}

\begin{acknowledgments}
The authors are indebted to Dr. H. T. Wong and Dr. M. Deniz of TEXONO
Collaboration for the detailed communications. A. N. K. would like to thank
The University of Kansas (KU) Particle Theory Group for hosting his visit
while this work was begun, the HEC of Pakistan for supporting his graduate
studies under the Indigenous Ph.D. Fellowship Program (Batch-IV). D.W. M.
and A. N. K received support from DOE Grant. No. De-FG02-04ER41308 and A. N.
K also recieved partial support from Saeeda-Iftikhar Scholarship award.
\end{acknowledgments}


\begin{thebibliography}{99}
\bibitem{nsifootnote} For example, the MINOS experiment has constrained the $%
\mu -\tau $ flavor-changing (FC) neutral-current (NC) parameter \cite%
{MINOSnsi}, which affects the propagation of $\nu _{\mu }$, while Ref. \cite%
{gmp} finds an allowed range of NSI parameters that allow the differing
central values for $\sin ^{2}(2\theta _{13})$ measured by \cite{T2K} and 
\cite{DB} to be compatible with each other, even if their central values
remain unchanged as their uncertainties shrink.
\bibitem{nocorrel} For cases where there is no systematic correlation
between parameters, the one parameter and two parameter bounds on the
individual parameters are the same anyway.
\end{thebibliography}
\end{document}